\pdfoutput=1
\documentclass[aps,prd,showpacs,preprintnumbers,twocolumn]{revtex4}

\usepackage[colorlinks=true, pdfstartview=FitV, linkcolor=blue, citecolor=blue, urlcolor=blue]{hyperref}
\usepackage[dvipdfmx]{graphicx}
\usepackage{amsmath}
\usepackage{color}

\newcommand{\tr}{\mathrm{tr}}
\newcommand{\Retr}{\mathrm{Retr}}

\begin{document}

\title{Lattice QCD in curved spacetimes}

\author{Arata~Yamamoto}
\affiliation{
Department of Physics, The University of Tokyo, Tokyo 113-0033, Japan\\
Theoretical Research Division, Nishina Center, RIKEN, Saitama 351-0198, Japan
}

\date{\today}

\begin{abstract}
We formulate the lattice QCD simulation with background classical gravitational fields.
This formulation enables us to study nonperturbative aspects of quantum phenomena in curved spacetimes from the first principles.
As the first application, we perform the simulation with the Friedmann-Lema\'itre-Robertson-Walker metric and analyze particle production in the expanding universe.
\end{abstract}


\maketitle

\section{Introduction}

Quantum field theory in curved spacetimes has a broad range of applicability.
It covers not only real gravitation, e.g., the expanding universe and black holes, but also non-inertial reference frames, e.g., accelerated frames and rotating frames.
One of the most prominent quantum phenomena in curved spacetimes is particle creation from the vacuum.
The particle creation occurs on black holes, in accelerated frames, and in the expanding universe \cite{Fulling:1972md}.
The understanding of such a quantum process at the full quantum level is a long standing problem.

Despite enormous efforts, the consistent quantization of gravity is still difficult due to nonrenormalizability.
At energies below the Planck scale, quantum effects of gravity can be neglected and gravity can be treated as a classical field.
Whereas quantum gravity is too difficult, quantum field theory with classical gravity is, at least in principle, tamable.
Practical calculations of interacting quantum field theory are, however, not easy even if gravity is classical,.
In particular, in the strong coupling region of quantum chromodynamics (QCD), a perturbative approach does not work successfully.
To study nonperturbative aspects of QCD, we need the lattice simulation, which is an {\it ab initio} nonperturbative approach in QCD.

In this paper, we formulate lattice QCD with external gravitational fields.
The gravitational fields are classical backgrounds in this framework, unlike in lattice quantum gravity \cite{Hamber:2009mt}.
While the backreaction from QCD to gravity is absent, quantum effects of QCD are exactly taken into account.
Lattice QCD in a curved spacetime was first formulated in a specific case of a rotating frame \cite{Yamamoto:2013zwa}.
We extend this formalism to general curved spacetimes.
There are pioneering works of Abelian gauge theory on curved lattices \cite{Jersak:1996mn}.
In addition, some kinds of simulations in flat spacetimes can be regarded as simulations in curved spacetimes.
The examples are an anisotropic lattice with direction-dependent coupling constant \cite{Hasenfratz:1981tw} and an inhomogeneous lattice with coordinate-dependent coupling constant \cite{Huang:1990jf}.

\section{Formulation}

We consider a four-dimensional Riemannian spacetime with the invariant length
\begin{equation}
 ds^2 = g_{\mu\nu}(x) dx^\mu dx^\nu.
\end{equation}
The metric tensor $g_{\mu\nu}(x)$ has positive signatures. 
For simplicity, we assume that the spacetime is covered with a single global coordinate patch.

In the continuum, the Yang-Mills action is
\begin{equation}
S_{\rm YM} = \int d^4x \sqrt{\det g} \ \frac{1}{2g_{\rm YM}^2} g^{\mu\nu} g^{\rho\sigma} \tr F_{\mu\rho} F_{\nu\sigma}.
\end{equation}
and the fermion action is
\begin{equation}
S_{\rm F} = \int d^4x \sqrt{\det g} \ \bar{\psi}[\gamma^\mu (\partial_\mu+iA_\mu+i\Gamma_\mu) +m] \psi
\end{equation}
\cite{Misner:1974qy}.
The connection is
\begin{equation}
\Gamma_\mu = \frac{1}{4}\sigma^{ij} \omega_{\mu ij} 
\end{equation}
with
\begin{eqnarray}
\sigma^{ij} &=& \frac{i}{2} [\gamma^i , \gamma^j]\\
\omega_{\mu ij} &=& g_{\alpha\beta} e^\alpha_i ( \partial_\mu e^\beta_j + \Gamma^\beta_{\mu\nu} e^\nu_j)\\
\Gamma^\beta_{\mu\nu} &=& \frac{1}{2} g^{\beta\rho} \left( \partial_\mu g_{\rho\nu} + \partial_\nu g_{\mu\rho} - \partial_\rho g_{\mu\nu} \right).
\end{eqnarray}
The Greek and Latin indices refer to the coordinate and tangent spaces, respectively.
They are related through the vierbein $e_i^\mu(x)$, which satisfies $e_i^\mu e_j^\nu g_{\mu\nu} = \delta_{ij}$.
The gamma matrix in curved spacetimes is given
\begin{equation}
\gamma^\mu(x) = \gamma^i e_i^\mu(x) ,
\end{equation}
where $\gamma^i$ is the gamma matrix in the flat Euclid space.
In general, there exists the ambiguity of the choice for the vierbein and the Dirac operator depends on it.

We embed a hypercubic lattice into a spacetime where the gravitational field and the coordinate are fixed.
Thus there is no general covariance, i.e., no local gauge invariance of gravity, unlike the dynamical triangulation in lattice quantum gravity \cite{Hamber:2009mt}.
On the lattice, continuum spacetime symmetry is broken to discrete symmetries.
For example, a hypercubic lattice has the symmetries of reflection, discrete rotation of $\pi/2$, and discrete translation of $a$.
Some of them are further broken by background gravitational fields.
The full spacetime symmetry depends on $g_{\mu\nu}(x)$.

We here consider the hypercubic lattice with a single lattice spacing $a$ that is independent of positions and directions, i.e.,
\begin{equation}
\int dx^\mu = a
\end{equation}
between nearest neighbor sites.
The $SU(N_c)$ link variable is obtained by discretizing the path-ordered product of $A_\mu dx^\mu$,
\begin{equation}
U_\mu(x) = \mathcal{P} e^{i\int dx^\mu A_\mu(x)} = e^{iaA_\mu(x)}.
\end{equation}
In this and the following equations, the contraction of the Lorentz indices is not performed unless otherwise explicitly summed.

The building block of the lattice gauge action is the plaquette
\begin{equation}
U_{\mu\nu}(x) = U_\mu(x) U_\nu(x+\hat{\mu}) U^\dagger_\mu(x+\hat{\nu}) U^\dagger_\nu(x) .
\end{equation}
The shorthand notation $\hat{\mu}$ means the unit lattice vector in the $x^\mu$ direction.
We construct three symmetric combinations of the plaquettes,
\begin{eqnarray}
\bar{U}_{\mu\nu} &=& \frac{1}{4} [ U_{\mu\nu} + U_{-\mu\nu} + U_{\mu-\nu} + U_{-\mu-\nu} ] \\
\bar{V}_{\mu\nu\rho} &=& \frac{1}{8} [ ( U_{\mu\nu} - U_{-\mu\nu} )( U_{\nu\sigma}-U_{\nu-\sigma} ) \nonumber \\
&& + ( U_{\mu-\nu} - U_{-\mu-\nu} )( U_{-\nu\sigma}-U_{-\nu-\sigma} ) ]  \\
\bar{W}_{\mu\nu\rho\sigma} &=& \frac{1}{16} [ U_{\mu\nu} - U_{-\mu\nu} - U_{\mu-\nu} + U_{-\mu-\nu} ] \nonumber \\
&& \times [ U_{\rho\sigma} - U_{-\rho\sigma} - U_{\rho-\sigma} + U_{-\rho-\sigma} ].
\end{eqnarray}
The plaquettes with negative directions are defined as $U_{\mu-\nu}(x) = U_\mu(x) U^\dagger_\nu(x+\hat{\mu}-\hat{\nu}) U^\dagger_\mu(x-\hat{\nu}) U_\nu(x-\hat{\nu})$ etc.
In the continuum limit,
\begin{eqnarray}
\Retr \bar{U}_{\mu\nu} &=& N_c - \frac{a^4}{2} \Retr F_{\mu\nu} F_{\mu\nu} \\
\Retr \bar{V}_{\mu\nu\rho} &=& - a^4 \Retr F_{\mu\nu} F_{\nu\rho} \\
\Retr \bar{W}_{\mu\nu\rho\sigma} &=& - a^4 \Retr F_{\mu\nu} F_{\rho\sigma}.
\end{eqnarray}
The lattice gauge action in curved spacetimes is
\begin{equation}
\begin{split}
S_{\rm YM} =& \frac{1}{g_{\rm YM}^2} \sum_{x} \sqrt{\det g(x)} \\
& \times\Bigg[ \sum_{\mu\ne\nu} g^{\mu\mu}(x) g^{\nu\nu}(x) \left( N_c -\Retr \bar{U}_{\mu\nu}(x) \right) \\
& - \frac{1}{2} \sum_{ \{\mu\nu\rho\} } g^{\mu\nu}(x) g^{\nu\rho}(x) \Retr \bar{V}_{\mu\nu\rho}(x) \\
& - \frac{1}{2} \sum_{ \{\mu\nu\rho\sigma\} } g^{\mu\rho}(x) g^{\nu\sigma}(x) \Retr \bar{W}_{\mu\nu\rho\sigma}(x) \Bigg].
\end{split}
\end{equation}
The summation $\sum_{ \{\mu\nu\rho\} }$ is performed so as to satisfy $\mu \ne \nu \ne \rho \ne \mu$ and the summation $\sum_{ \{\mu\nu\rho\sigma\} }$ is performed such that $(\mu\nu\rho\sigma)$ is a permutation of $(1234)$.
In this construction, we adopted the condition that it respects reflection symmetry (apart from explicit symmetry breaking by the metric) and has the smallest number of loops.
As long as the continuum limit is the same, other constructions are possible.
For example, $\bar{V}_{\mu\nu\rho}$ can be replaced by $\bar{W}_{\mu\nu\nu\rho}$ but it has larger number of loops.

Among several choices of lattice fermions, we here consider the simplest one, i.e., the Wilson fermion.
The Wilson fermion action in curved spacetimes is
\begin{equation}
\begin{split}
S_{\rm F} =& \sum_{x_1 x_2} a^3 \bar{\psi}(x_1) \Big[ (am+4) \sqrt{\det g(x_1)}\delta_{x_1,x_2} \\
& - \frac{1}{2} \sum_\mu \big\{(1-\gamma^\mu(x_1)) \sqrt{\det g(x_1)} \\
& \times  V_\mu(x_1) U_{\mu}(x_1) \delta_{x_1+\hat{\mu},x_2} + (1+\gamma^\mu(x_2)) \\
& \times \sqrt{\det g(x_2)} V^\dagger_\mu(x_2) U^\dagger_{\mu}(x_2)\delta_{x_1-\hat{\mu},x_2} \big\} \Big] \psi(x_2) .
\end{split}
\label{eqSflat}
\end{equation}
The connection is introduced as the Spin(4) link variable
\begin{equation}
V_\mu(x) = e^{ia\Gamma_\mu(x)} .
\end{equation}
The splitting of the arguments ($x_1$ or $x_2$) in Eq.~(\ref{eqSflat}) between adjacent lattice sites is determined by requiring $\gamma^5$ Hermiticity of the lattice Wilson-Dirac operator $\gamma^5 D \gamma^5 = D^\dagger$.
In the formalism of the Wilson fermion, the so-called Wilson term is added to the naive fermion action to kill the artificial poles of doublers.
Since the detailed form of the Wilson term is irrelevant in the continuum limit, we added the usual Wilson term which is the Laplacian $\Delta_\delta = \delta^{\mu\nu} \partial_\mu \partial_\nu$ in a flat spacetime.
We can use the Laplacian $\Delta_g = (1/\sqrt{\det g})\partial_\mu g^{\mu\nu} \sqrt{\det g} \partial_\nu$ although the lattice action becomes more complicated.

As in lattice QCD in flat spacetimes, improved lattice actions can be constructed by adding higher-order terms of the lattice spacing, which are irrelevant in the continuum limit, to reduce discretization artifacts.

\section{Wick rotation}

In the Euclidean path integral formulation, real time $t$ is transformed to imaginary time $\tau$ by the Wick rotation $\tau = -it$.
In Euclidean quantum gravity, the Wick rotation causes serious problems, such as the conformal instability \cite{Gibbons:1976ue}.
Most of the problems are irrelevant for classical gravity because they originate from the dynamical Einstein-Hilbert action and the diffeomorphism invariance.
However, also in classical gravity, the Wick rotation causes the complex metric problem.

For instance, if $g_{\mu 0}(t)$ ($\mu \ne 0$) is nonzero or if $g_{\mu \nu}(t)$ includes an odd function of $t$, then $g_{\mu\nu}(\tau)$ is complex and thus the Euclidean action is complex.
Also, if the connection is an imaginary number, the fermion action is complex.
Even if all the elements of the metric tensor is real, as in the Minkowski space, the action can be negative and thus non-positive definite.
When the total action is not positive definite, the Monte Carlo simulation does not work because of the sign fluctuation.
This is called the sign problem.
This is a known problem of a quark chemical potential and an external electric field \cite{deForcrand:2010ys}.
The fermion action with a chemical potential or an electric field is complex.
In curved spacetimes, the sign problem is more severe because both of the gauge and fermion actions can be complex.

One approach to avoid the sign problem is to change the real parameter that makes the action complex to the imaginary parameter.
This is the analogy of an imaginary quark chemical potential and a Euclidean electric field \cite{deForcrand:2010ys}.
The real information in the Lorentzian spacetime is obtained by analytic continuation.
Thus this approach is justified only when analytic continuation is validated.
If analyticity is lost, for example in the presence of a phase transition, this approach is not justified.
In numerical simulations, analytic continuation is done as numerical extrapolation along the parameter.
This extrapolation is expected to be reliable for small parameter region, i.e., in weakly curved spacetimes.

Another approach is to utilize special symmetry to cancel the complex phase of the action.
For the sign problem of chemical potentials and electric fields, there are several known symmetries, e.g., isospin symmetry \cite{Son:2000xc}.
For the complex metric problem, such symmetry is not yet known.
In contrast to the first approach, the second approach is applicable to large parameter region, i.e., in strongly curved spacetimes.

An entirely different approach is stochastic quantization \cite{Damgaard:1987rr}.
Although stochastic quantization for complex action is not fully understood, it is developing rapidly.
The direct simulation of complex metric systems might be possible in the future.

\section{Renormalization}

The ultraviolet divergence of quantum field theory comes from infinitely short length scale.
When the gravitational field is classical, it has only fixed intrinsic scales.
Therefore classical gravity does not cause new ultraviolet divergences.
Actually, the continuum theory is known to be renormalizable in general curved spacetimes \cite{Buchbinder:1989zz}.
Although there is no formal proof of the renormalizability by lattice regularization, the lattice theory is expected to be renormalizable.

The lattice spacing is affected by renormalization.
The renormalization of the lattice spacing is troublesome in curved spacetimes.
In flat spacetimes, if the classical lattice spacing is homogeneous and isotropic, the renormalized lattice spacing is homogeneous and isotropic because it is protected by spacetime symmetry.
In the curved spacetimes where spacetime symmetry is explicitly broken, even if the classical lattice spacing is homogeneous and isotropic, the renormalized lattice spacing can be inhomogeneous and anisotropic.
A famous example is the renormalization of the anisotropic lattice action \cite{Hasenfratz:1981tw}.
On the anisotropic lattice, the spatial lattice spacing $a_s$ and the temporal lattice spacing $a_\tau$ are different.
Since they are differently affected by the renormalization, the renormalized anisotropic ratio $\xi_{\rm ren}$ deviates from the classical value $\xi_{\rm cl}=a_s/a_\tau$.
In weakly curved spacetimes, the renormalization correction may approximately be neglected.
In a strongly curved spacetime, we need to find a (perturbative or nonperturbative) scheme to determine the renormalized lattice spacing and the physical unit.

In addition to the quantum correction, the lattice spacing receives a classical gravitational correction.
In a curved spacetime, the invariant length is $ds = \sqrt{g_{\mu\nu} dx^\mu dx^\nu}$.
The distance between nearest neighbor sites changes as $a \to \sqrt{g_{\mu\mu}}a$.
When we calculate a two-point correlator as a function of time or distance, we should use the proper time or the proper length $l = \int ds = \sum \sqrt{g_{\mu\mu}}a$.

\section{Simulation}

We explicitly demonstrate the computational implementation of the above formulation.
For a simple and intuitive example, we consider particle production in an expanding space.
The time evolution of a flat three-dimensional space is described by the Friedmann-Lema\'itre-Robertson-Walker metric
\begin{equation}
 ds^2 = d\tau^2 + \alpha(\tau)^2 (dx^2 + dy^2 + dz^2),
\end{equation}
which is used for the cosmological model of the expanding universe \cite{Friedmann:1924bb}.
The functional form of the scale factor $\alpha(\tau)$ depends on the contents of the universe, e.g., nonrelativistic matter, radiation, or the cosmological constant.

The lattice gauge action is
\begin{equation}
\begin{split}
S_{\rm YM} =& \beta \sum_{x} \Bigg[ \sum_{k} \alpha(\tau) \left( 1 - \frac{1}{N_c}\Retr \bar{U}_{4k}(x) \right) \\
&+ \sum_{k > l} \alpha(\tau)^{-1} \left( 1 - \frac{1}{N_c}\Retr \bar{U}_{kl}(x) \right) \Bigg]
\end{split}
\label{eqSFLRW1}
\end{equation}
with $k,l=1,$ 2, and 3.
The lattice coupling constant is defined as $\beta = 2N_c/g_{\rm YM}^2$.
This action is similar to the anisotropic lattice action \cite{Hasenfratz:1981tw} but the coefficient depends on coordinates.
The lattice fermion action is
\begin{equation}
\begin{split}
S_{\rm F} =& \sum_{x_1 x_2} a^3 \bar{\psi}'(x_1) \Big[ \delta_{x_1,x_2} \\
&- \kappa \sum_k \big\{(1-\gamma^k\alpha(\tau_1)^{-1}) V_k (x_1) U_k (x_1) \delta_{x_1+\hat{k},x_2} \\
&+ (1+\gamma^k\alpha(\tau_2)^{-1}) V^\dagger_k (x_2) U^\dagger_k (x_2) \delta_{x_1-\hat{k},x_2} \big\} \\
&- \kappa \big\{(1-\gamma^4) \left( \frac{\alpha(\tau_1)}{\alpha(\tau_2)} \right)^{\frac{3}{2}} U_{4}(x_1) \delta_{x_1+\hat{4},x_2} \\
&+ (1+\gamma^4) \left( \frac{\alpha(\tau_2)}{\alpha(\tau_1)} \right)^{\frac{3}{2}} U^\dagger_{4}(x_2)\delta_{x_1-\hat{4},x_2} \big\} \Big] \psi'(x_2).
\end{split}
\label{eqSFLRW2}
\end{equation}
The spinor fields are rescaled as $\bar{\psi}'(x_1) = \alpha(\tau_1)^{3/2}(am+4)^{1/2}\bar{\psi}(x_1)$ and $\psi'(x_2) = \alpha(\tau_2)^{3/2}(am+4)^{1/2}\psi(x_2)$.
The hopping parameter is defined as $\kappa = 1/(2am+8)$.
The connection is
\begin{equation}
V_k(x) = \exp \left(i \gamma^k \gamma^4 \frac{\partial_4 \alpha(\tau)}{2} \right).
\end{equation}
The vierbein is taken as $e_1^1 = e_2^2 = e_3^3 = 1/\alpha$, $e_4^4 = 1$,
and $ e_i^\mu = 0$ for $\mu \ne i$.

In this study, we consider the expanding universe with the cosmological constant.
The scale factor is $\alpha(t) = \alpha_0 e^{Ht}$ in the Lorentzian spacetime.
The parameter $H$ is the Hubble constant.
The naive Wick rotation to imaginary time provides the complex scale factor $\alpha(\tau) = \alpha_0 e^{iH\tau}$, and causes the sign problem.
To avoid this, we introduce the ``imaginary'' Hubble constant $H_I=iH$ and the Euclidean expansion as
\begin{equation}
\alpha(\tau) = \alpha_0 e^{H_I\tau}.
\end{equation}
The metric tensor is real and the lattice action is positive definite, as seen in Eqs.~(\ref{eqSFLRW1}) and (\ref{eqSFLRW2}).
Note that we cannot directly relate the following result to particle production in the Lorentzian spacetime without analytic continuation.
We here treat the Euclidean expansion itself, as the theoretical study of QCD with an imaginary chemical potential.

\begin{figure}[h]
\includegraphics[scale=1]{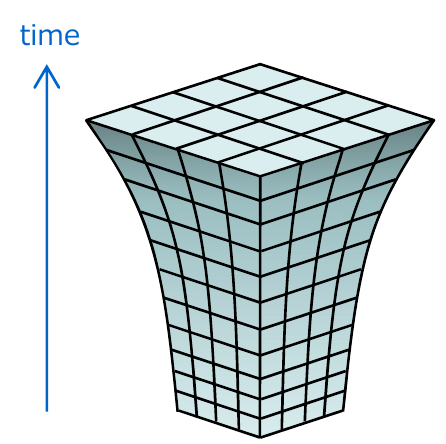}
\caption{\label{fig1}
Expanding lattice.
}
\end{figure}

The geometry is schematically shown in Fig.~\ref{fig1}.
The three-dimensional space starts to expand at $\tau=0$ and ends at $\tau=a(L_\tau-1)$, where Dirichlet boundary conditions are imposed.
The initial scale factor is set to $\alpha(0)=\alpha_0=1$.
In the three-dimensional space, periodic boundary conditions are imposed.
The lattice size is $L_xL_yL_z \times L_\tau = 10^3 \times 20$.
We performed quenched QCD simulation with $\beta = 5.9$ and $\kappa = 0.154$.
We only consider small parameter region $aH_I \ll 1$.

\begin{figure}[h]
\includegraphics[scale=1.2]{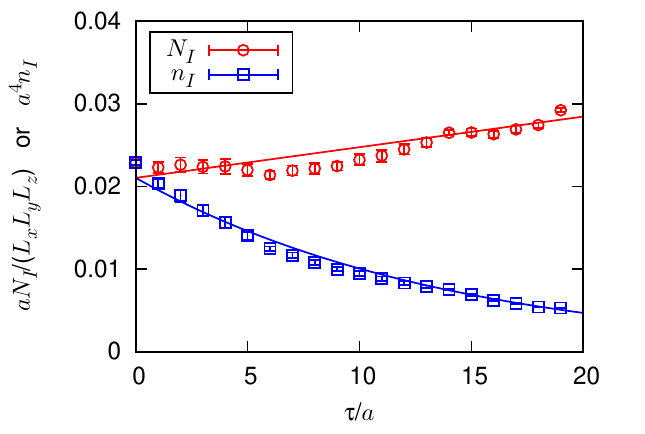}
\caption{\label{fig2}
$\tau$-dependence of the total fermion number $N_I$ and the fermion number density $n_I$ with $aH_I=0.03$.
}

\includegraphics[scale=1.2]{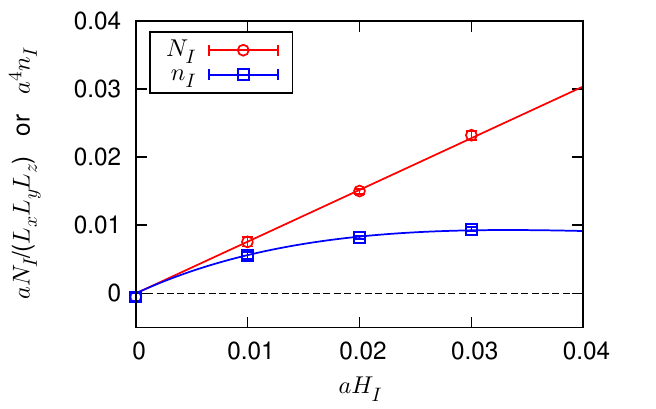}
\caption{\label{fig3}
$H_I$-dependence of the total fermion number $N_I$ and the fermion number density $n_I$ at $\tau/a=10$.
}
\end{figure}

For particle production in the Euclidean expansion, we computed the imaginary particle number of fermions at fixed time slices
\begin{eqnarray}
N_I(\tau) &=& \int d^3x \sqrt{\det g}\ n_I (x) \\
n_I(x) &=& -i j^4(x) = -i \langle \bar{\psi}(x) \gamma^4 \psi(x) \rangle .
\end{eqnarray}
As shown in Figs.~\ref{fig2} and \ref{fig3}, nonzero positive $N_I$ and $n_I$ are produced in the Euclidean expansion.
The data of $N_I$ is rescaled by multiplying a factor $1/(L_xL_yL_z) = 10^{-3}$.
The inequality $N_I/(L_xL_yL_z) \ge n_I$ holds in this expanding space because of $\sqrt{\det g}=\alpha^3 \ge 1$.
From numerical fitting, we obtained $N_I/(L_xL_yL_z) = C_1 H_I [(\tau/a) + C_2]$ and $n_I = C_1 H_I [(\tau/a) + C_2] \alpha^{-3}$ with $C_1 = 0.012 \pm 0.001$ and $C_2 = 57 \pm 8$.
The best-fit functions are shown in the figures.

\section{Conclusion}

We have formulated lattice QCD in curved spacetimes to study gravitational effects on QCD.
We have performed the first simulation to demonstrate the computational implementation.
For practical applications to phenomenology, there are open issues to be discussed in more detail, in particular, analytic continuation and renormalization.
The full QCD simulation is also necessary for the study of particle production.
By performing quantitative analysis, we can discuss nonperturbative QCD effects on cosmology.
For example, the functional dependence on the Hubble constant is essential for a scenario of dark energy \cite{Zhitnitsky:2013pna}.

There are a large number of future developments of this framework.
On the theoretical side, we can also formulate other kinds of lattice field theory in curved spacetimes, e.g., scalar field theory, electroweak gauge theory, and nonrelativistic field theory, and so on.
In scalar field theory, the action includes the renormalizable term $R \phi^2$, which couples to a scalar curvature $R$.
On the practical side, by applying this framework, we can study nonperturbative phenomena of QCD in various curved spacetimes, e.g., on black holes, in the anti-de Sitter space, and so on.

We have considered only the case that the spacetime is covered with a single regular coordinate patch.
In several physically interesting spacetimes, the metric tensor $g_{\mu\nu}$ is singular, e.g., on black holes, or the inverse metric tensor $g^{\mu\nu}$ is singular, e.g., in polar coordinates.
When such singularities exist, we need the scheme to resolve it: (i) transforming a singular coordinate to a regular one, (ii) cutting the region around the singularities, or (iii) introducing several local coordinate patches and gluing them at boundary regions.
The formulation of this scheme on the lattice is also a future work.

\acknowledgments
The author is grateful to Kenji Fukushima and Yuya Tanizaki for useful discussions.
The numerical simulations were performed by using the RIKEN Integrated Cluster of Clusters (RICC) facility.

\end{document}